\documentstyle[11pt,amstex,righttag]{article}
\newcommand{\bx}{\text{\bf x}}
\newcommand{\dd}{\text{d}}
\newcommand{\ee}{\text{e}}

\renewcommand{\text}{\mbox}

\begin{document}

\thispagestyle{empty}
\title{Persistence exponent of the diffusion equation\\ 
in $\epsilon$ dimensions\\
${}$}

\author{H.J. Hilhorst\\
Laboratoire de Physique Th\'eorique$^1$\\
B\^atiment 210\\
Universit\'e de Paris-Sud\\
91405 Orsay cedex, France\\}

\maketitle
\vspace{-1cm}
\begin{small}
\begin{abstract}
\noindent 
We consider the $d$-dimensional diffusion equation
$\partial_t\phi(\bx,t)=\Delta\phi(\bx,t)$ with random initial
condition, and
observe that, when appropriately scaled, $\phi(0,t)$ is Gaussian
and Markovian in the limit $d\to 0$. This leads via the Majumdar-Sire
perturbation theory to a small $d$ expansion for the persistence exponent
$\theta(d)$. We find $\theta(d)=\frac{1}{4}d-0.12065$...$
d^{3/2}+\ldots.$
\end{abstract}
\end{small}
\vspace{87mm}

\noindent 
LPT ORSAY 99/71\\
\noindent
{\small$^1$Laboratoire associ\'e au Centre National de la
Recherche Scientifique - UMR 8627}
\newpage

\noindent
We consider the $d$-dimensional linear
diffusion equation 
\begin{equation}
\frac{\partial\phi(\bx,t)}{\partial t}=\Delta\phi(\bx,t)
\label{one}
\end{equation}
with initial condition $\phi(\bx,0)=\psi(\bx)$, where $\psi(\bx)$ is a
zero mean Gaussian random field of covariance
$\langle\psi(\bx)\psi(\bx')\rangle=\delta(\bx-\bx')$.
We are interested in the persistence of the stochsastic process
$\Phi_d(t)=\phi(0,t),$ that is, in the $t\to\infty$\,\, limit of the
probability $Q(t)$ that in the interval $0\leq t'\leq t$ the process
$\Phi_d(t)$ does not change sign. This question was first asked by
Majumdar {\it et al.} \cite{MSBC} and Derrida {\it et al.}
\cite{DHZ}. The asymptotic decay of $Q(t)$ appears to be 
given by a power law
\begin{equation}
Q(t)\sim t^{-\theta(d)}
\label{two}
\end{equation} 
with a {\it persistence exponent} $\theta(d)$ whose precise value has
been the main focus of study. 
One particular application \cite{DHZ} 
concerns the coarsening of a phase-ordering
system with nonconserved order parameter \cite{OJK}.

Persistence exponents of random processes, 
also of interest in many
other contexts in physics
(see a recent review due to Majumdar \cite{Majumdar}), have
turned out to be very hard to calculate. A perturbative method in this
field was designed by Majumdar and Sire \cite{MajumdarSire} and
simplified by Oerding {\it et al.} \cite{OCB} (see also Majumdar {\it et
al.} \cite{SMR}). It consists of expanding around a Gaussian Markovian
process, for which the persistence exponent is
known. This method was applied to zero temperature \cite{MajumdarSire}
and critical point \cite{OCB} coarsening in the
Ising model. 
More recently Majumdar and Bray
\cite{MajumdarBray} showed how to expand
the persistence exponent of
a Gaussian process in powers of a parameter $1\!-\!p$
interpretable as a "fugacity"
weighting the sign changes of the random process.

For the diffusion equation (\ref{one}) 
no exact 
expressions for
$\theta(d)$ are known. Approximate numerical values
come from simulation data
\cite{MSBC}, from
the "independent interval approximation" (IIA) \cite{MSBC,DHZ},
and from work by Newman and Toroczkai \cite{NewmanToroczkai}
based on a
hypothesis concerning the "sign time distribution" 
(see also Dornic and Godr\`eche \cite{DornicGodreche} and Drouffe and
Godr\`eche \cite{DrouffeGodreche}).  
In this note we
show that the persistence exponent $\theta(d)$ allows for an
$\epsilon$ expansion in dimension $0+\epsilon$.\\

Using the Green function of the diffusion equation and writing $S_d$ for
the surface of the $d$-dimensional unit sphere we can solve (\ref{one})
and find
\begin{equation}
\Phi_d(t)=\frac{S_d^{1/2}}{(4\pi t)^{d/2}}\,\int_0^\infty\!\dd r\,
r^{\frac{1}{2}(d-1)}\, \ee^{-r^2/4t}\,\Psi(r)
\label{three}
\end{equation}
where $\Psi(r)$ is the appropriately normalized integral of the initial
field $\psi(\bx)$ on a spherical shell of radius $r$ around the origin,
\begin{equation}
\Psi(r)=S_d^{-1/2}r^{-\frac{1}{2}(d-1)}\lim_{\Delta r\to 0}
\frac{1}{\Delta r}\,\int_{r<x<r+\Delta r}\!\dd\bx \,\psi(\bx)
\label{four}
\end{equation}
Hence $\Psi(r)$ is Gaussian of mean zero and covariance
$\langle\Psi(r)\Psi(r')\rangle=\delta(r-r').$ 
We shall henceforth write $d=\epsilon$
to indicate that the RHS of
(\ref{three}) is continued
analytically to noninteger dimensions and that we prepare
for the limit of zero dimension.

The remaining discussion is easiest in terms of the new time variable
$\tau=\epsilon\log t$ and the rescaled process
\begin{equation}
\tilde{\Phi}_\epsilon(\tau)=(8\pi)^{\epsilon/4}\,\ee^{\tau/4}\,
\Phi_\epsilon
(\ee^{\tau/4\epsilon})  
\label{five}
\end{equation}
which has the advantage of being stationary. 
Let us denote by
$\tilde{Q}(\tau)$ the persistence probability 
of $\tilde{\Phi}_\epsilon(\tau)$. This quantity then decays as
$\tilde{Q}(\tau)\sim\exp(-\tilde{\theta}(\epsilon)\tau)$ with 
$\epsilon\tilde{\theta}=\theta.$ We consider now the 
explicit expression for the correlator
$F_\epsilon(\tau-\tau')\equiv
\langle\tilde{\Phi}_\epsilon(\tau)\tilde{\Phi}_\epsilon(\tau)\rangle,$
which reads
\begin{equation}
F_\epsilon(\tau-\tau')=\Big(\cosh\frac{\tau-\tau'}{2\epsilon}\Big)
^{-\epsilon/2}
\label{six}
\end{equation}
We observe that in the limit $\epsilon\to 0$ we have the exponential
decay $F_0(\tau-\tau')=\exp(-\tilde{\lambda}|\tau-\tau'|)$
with $\tilde{\lambda}=\frac{1}{4}.$
It follows that $\tilde{\Phi}_0(t)$ is Markovian and that it has
$\,\tilde{\theta}(0)=\tilde{\lambda}\,$ (see {\it e.g.} \cite{Majumdar}),
whence we conclude that $\theta(0)=0.$

For $\epsilon>0$ the process $\tilde{\Phi}_\epsilon(\tau)$ is
non-Markovian and there is no known way
to calculate $\tilde{\theta}(\epsilon)$
exactly. We therefore consider $F_\epsilon-F_0$ as a
small perturbation. This allows us to apply the principal result of the
Majumdar-Sire perturbation theory \cite{MajumdarSire}. Cast in a form
first explicitly exhibited by Oerding {\it et al.} \cite{OCB} 
and applied to the
problem at hand it states that to lowest order in $F_\epsilon-F_0$ one
has
\begin{equation}
\tilde{\theta}(\epsilon) = \tilde{\lambda}\,\Big[\,1\,-\,
\frac{2\tilde{\lambda}}{\pi}\int_0^\infty\!\dd\tau\,\,\frac
{F_\epsilon(\tau)-F_0(\tau)}{(1-\ee^{-2\tilde{\lambda}\tau})^{3/2}}\,\Big]
\label{seven}
\end{equation}
Upon substituting in (\ref{seven}) the explicit expressions for
$\tilde{\lambda},\, F_\epsilon,$ and $F_0$, and multiplying by
$\epsilon$, we get
\begin{equation}
\theta(\epsilon)=\frac{\epsilon}{4}\,\Big[\,1\,-\,
\frac{1}{2\pi}\int_0^\infty\!\dd\tau\,\,
\frac{(\cosh\frac{\tau}{2\epsilon})^{-\epsilon/2}-\ee^{-\tau/4}}
{(1-\ee^{-\tau/2})^{3/2}}\,\Big]
\label{eight}
\end{equation}
We extract from (\ref{eight})
the first two terms of the $\epsilon$ expansion by setting $\tau=\epsilon
u$, expanding the integrand for small $\epsilon$, and performing an
integration by parts. The result is
\begin{eqnarray}
\theta(\epsilon)&=&\frac{\epsilon}{4}\,-\,
\frac{1}{\pi}\Big(\frac{\epsilon}{2}\Big)^{3/2}
\int_0^\infty\frac{\dd u}{\sqrt{u}}\,\,\frac{1}{1+\ee^u}\,+\,\ldots
\nonumber\\[2mm]
&=&\frac{\epsilon}{4} \,-\, (8\pi)^{-1/2}(1-\sqrt{2})
\,\,\zeta({\text {$\frac{1}{2}$}})
\,\,\epsilon^{3/2}\,+\,\ldots\nonumber\\[2mm]
&=&\frac{\epsilon}{4} \,-\, 0.12065\mbox{...}\,\epsilon^{3/2}\,+\,\ldots
\label{nine}
\end{eqnarray}
Truncating this series after the $\epsilon^{3/2}$ term we find 
$\theta(1)=0.1293$...\,, which should be compared to the
Monte Carlo 
estimate $\theta(1)=0.1207\pm0.0005$ \,\cite{MSBC},
to the IIA result
$\theta(1)=0.1203$...\,
\cite{MSBC,DHZ}, and to the value $\theta(1)=0.1253$... of
Ref.\,\cite{NewmanToroczkai}.
No numerical values for $\theta(1)$ are available
from the $1\!-\!p$ expansion of Ref.\,\cite{MajumdarBray}. 
Finally, retaining in (\ref{eight}) the full integral -- which is 
the expansion proposed by the original authors 
\cite{MajumdarSire} -- gives for $\epsilon=1$ the somewhat improved result
$\theta(1)=0.1267$...\,.\\

We conclude that the $\epsilon$ expansion (\ref{nine}) 
is significant as one of the rare
analytical results that exist today  
for the persistence exponent of the diffusion equation. To make it
numerically competitive higher orders in $\epsilon$ should be
calculated.



\begin{thebibliography}{10}

\bibitem{MSBC}
S.N. Majumdar, C. Sire, A.J. Bray, and S.J. Cornell,
\newblock {\it Phys. Rev. Lett.} {\bf 77} (1996) 2867.

\bibitem{DHZ}
B. Derrida, V. Hakim, and R. Zeitak,
\newblock {\it Phys. Rev. Lett.} {\bf 77} (1996) 2871.

\bibitem{OJK}
T. Ohta, D. Jasnow, and K. Kawasaki,
\newblock {\it Phys. Rev. Lett.} {\bf 49} (1982) 1223.

\bibitem{Majumdar}
S.N. Majumdar,
\newblock {\it preprint cond-mat/9907407}.

\bibitem{MajumdarSire}
S.N. Majumdar and C. Sire,
\newblock {\it Phys. Rev. Lett.} {\bf 77} (1996) 1420.

\bibitem{OCB}
K. Oerding, S.J. Cornell, and A.J. Bray,
\newblock {\it Phys. Rev. E} {\bf 56} (1997) R25.

\bibitem{SMR}
C. Sire, S.~N. Majumdar, and A. R\"udinger,
\newblock {\it preprint cond-mat/9810136}.

\bibitem{MajumdarBray}
S.N. Majumdar and A.J. Bray,
\newblock {\it Phys. Rev. Lett.} {\bf 81} (1998) 2626.

\bibitem{NewmanToroczkai}
T.J. Newman and Z. Toroczkai,
{\it Phys. Rev. E} {\bf 58} (1998) R2685.

\bibitem{DornicGodreche}
I. Dornic and C. Godr\`eche,
\newblock {\it J. Phys. A} {\bf 31} (1998) 5413.

\bibitem{DrouffeGodreche}
J.-M. Drouffe and C. Godr\`eche, 
\newblock {\it J. Phys. A} {\bf 31} (1998) 9801. 


\end{thebibliography}
\end{document}